\documentstyle[11pt,a4]{article}
\author{Emmanuel Kohlprath\footnote{e-mail address:
 Emmanuel.Kohlprath@cern.ch}\\
Theoretical Physics Division, CERN \\ CH - 1211 Geneva 23}
\title{{\normalsize\begin{flushright}
{CERN-TH/2002-146}\\
hep-th/0207023\\
\end{flushright}}\vspace*{1cm}
Renormalization of the Planck mass\\ for type II
 superstrings on symmetric orbifolds}

\newcommand{\be}{\begin{equation}}
\newcommand{\ee}{\end{equation}}
\newcommand{\ba}{\begin{eqnarray}}
\newcommand{\ea}{\end{eqnarray}}
\newcommand{\no}{\nonumber}

\date{}

\begin{document}
\maketitle

\begin{abstract}
We compute the one-loop renormalization of the Planck mass for type~II string
theories compactified to four dimensions on symmetric orbifolds that preserve
${\cal N}=2$ supersymmetry. Depending on the orbifold, the effect can be  
as large as to compete with the standard tree-level value. 
\end{abstract}

\section{Introduction}
The weakness of the gravitational force can be rephrased as an enormous
 hierarchy between the Planck mass $M_{\rm P}$ and the scales characterizing
 non-gravitational interactions. In the context of supersymmetric
Grand Unified Theories the hierarchy between the Planck scale $M_{\rm P}$ and
 the GUT scale $M_{\rm \small{GUT}}$ appears to be less dramatic (a mere
 $10^3$), but is nevertheless puzzling.

The natural framework for trying to understand this ratio is obviously
superstring theory, since it supposedly unifies all interactions. Such
 a theory contains a scale of its own, the string mass $M_{\rm s}$, 
which plays the role of a fundamental UV cut-off related to the finite size
 of quantum strings. It is quite natural to identify $M_{\rm s}$ with
 $M_{\rm \small{GUT}}$ (though not an absolute theoretical necessity) and, in
 this case, explaining the origin of the ratio $M_{\rm s}/M_{\rm P}$ becomes
 the problem.

Unfortunately, for closed strings and at tree level (or in  string-loop
 perturbation theory), this ratio gets related to
 $\sqrt{\alpha_{\rm \small{GUT}}}$ in such
 a way that, when all factors are included, one finds
 $M_{\rm s} \approx 2\times10^{17}$ GeV, i.e. an order of magnitude too large
 a value for identification with $M_{\rm \small{GUT}} \approx 2\times10^{16}$
 GeV. If we wish to insist on $M_{\rm s} \approx M_{\rm \small{GUT}}$,
 this leaves e.g. the possibilities of either considering open string
 theory, or going beyond the validity of the loop expansion (for other
 possibilities see \cite{Dienes96}).

A suggestion in the latter direction was recently made in \cite{GV}, where it
 was proposed that, in the infinite bare coupling (also known as
 compositeness) limit, loop corrections may explain $M_{\rm s}/M_{\rm P}
 \ll 1$ as a
 result of the different large-$N$ behaviour of gravity and gauge loops in a
 theory with a large number $N$ of particle species. The arguments given in
 \cite{GV} were quite heuristic since they assumed some UV completion of a toy
 model, making it UV-finite, as well as the existence of an arbitrarily large
 parameter $N$. In this paper we shall explore this suggestion (at one loop)
 in the context of superstring theories with ${\cal N}=2$ space-time
 supersymmetry. A larger value of ${\cal N}$ would presumably induce no
 one-loop renormalization of $M_{\rm P}$, while the non-supersymmetric case,
 ${\cal N}=0$, would be generally plagued by a large one-loop cosmological
 constant. Unfortunately, the models for which we shall be able to perform
 the calculation do not contain an arbitrary parameter like $N$. At the end
 of the paper we shall introduce models that do have this feature, but for
 which, at present, we are unable to carry out a full string-theory
 calculation.

\section{Background field method}\label{Backgroundfieldmethod}

The Einstein--Hilbert Lagrangian reads (we use units in which $\hbar=c=1$)
\be
{\cal L}=\frac{1}{16\pi G}\sqrt{g}R=\frac{M_{\rm P}^2}{16\pi}\sqrt{g}R.
\label{EinsteinHilbert}
\ee
In \cite{KiritsisKounnas95} Kiritsis and Kounnas compute the one-loop
 renormalization of the coupling of the Einstein--Hilbert action (i.e. of
 Newton's constant $G$ or of the Planck mass $M_{\rm P}$) for the
 compactification of type II string theory (both IIA and IIB) on the symmetric
 ${\bf Z}_2\times {\bf Z}_2$ orbifold to 4 dimensions. First we notice
 that this compactification preserves ${\cal N}=2$ space-time supersymmetry
 in 4 dimensions. If there were more unbroken supersymmetries, then there
 should be no one-loop correction to the Planck mass due to extra zero modes.
 We generalize their results in \cite{KiritsisKounnas95} using the background
 field method (see also \cite{KiritsisKounnas94} to
 \cite{KiritsisKounnasPetropoulosRizos96a}) to a large class of orbifold
 compactifications with ${\cal N}=2$ space-time supersymmetry in
 4 dimensions. We can choose the axes such that the point group generator
 for ${\bf Z}_N$ orbifolds has the form (see \cite{Polchinski} and
 \cite{BailinLove99} for the theory of orbifolds)
\be
\theta=\exp[2\pi i(v_2 J_{45} +v_3 J_{67} +v_4 J_{89})].
\ee
The criterion for space-time supersymmetry is then
\be
\pm v_2\pm v_3\pm v_4=0,\label{supersymmetryinorbifolds1}
\ee
for some choice of the signs.
If the $v_i$ are otherwise non-vanishing then we have ${\cal N}=2$ space-time
 supersymmetry. If one of the $v_i$'s is zero, then we have ${\cal N}=4$
 space-time supersymmetry. For the ${\bf Z}_N\times{\bf Z}_M$ orbifolds
 we have a second set of point-group generators
\be
\varphi=\exp[2\pi i(w_2 J_{45} +w_3 J_{67} +w_4 J_{89})].
\ee
The criterion for space-time supersymmetry then is
\be
\pm v_2\pm v_3\pm v_4=0,\qquad \pm w_2\pm w_3\pm w_4=0,
\label{supersymmetryinorbifolds2}
\ee
for some choice of the signs. Only if  $v_i=w_i=0$ for the same $i=2,3,4$ do
 we have ${\cal N}\geq 4$ space-time supersymmetry.
The requirement that the point group acts crystallographically and the
 conditions (\ref{supersymmetryinorbifolds1}) and
 (\ref{supersymmetryinorbifolds2}) lead to the orbifold models
 listed in tables \ref{ZNorbiflods} and \ref{ZNxZMorbiflods} (see e.g.
 \cite{BailinLove99}).
\begin{table}\begin{center}\begin{tabular}{lll}\hline 
${\rm Point\ group}$& $(v_2,v_3,v_4)$&$s$\\\hline 
${\bf Z}_3$& $(1,1,-2)/3$&8\\
${\bf Z}_4$& $(1,1,-2)/4$&12\\
${\bf Z}_6-I$& $(1,1,-2)/6$&32\\
${\bf Z}_6-II$& $(1,2,-3)/6$&24\\
${\bf Z}_7$& $(1,2,-3)/7$&48\\
${\bf Z}_8-I$& $(1,2,-3)/8$&60\\
${\bf Z}_8-II$& $(1,3,-4)/8$&48\\
${\bf Z}_{12}-I$& $(1,4,-5)/12$&128\\
${\bf Z}_{12}-II$& $(1,5,-6)/12$&108\\
\hline\end{tabular}\end{center}\caption{Point-group generators for
 ${\bf Z}_N$ orbifolds}\label{ZNorbiflods}\end{table}
Let us define the complex combinations
\be
Z^i=\frac{1}{\sqrt{2}}(X^{2i}+iX^{2i+1}),\qquad i=2,3,4
\ee
for the compactified bosonic coordinates and
\be
\Psi^i=\frac{1}{\sqrt{2}}(\psi^{2i}+i\psi^{2i+1}),\qquad i=2,3,4
\ee
for the fermions. The orbifold twist acts as
\be
Z^i(\sigma+2\pi)=e^{2\pi iv_i}Z^i(\sigma)
\ee
on the bosons and as
\be
\tilde\Psi^i(\sigma+2\pi)=e^{2\pi i(v_i+\nu)}\tilde\Psi^i(\sigma)
\ee
on the fermions, where $\nu=0$ in the Ramond sector and $\nu=1/2$ in the
 Neveu--Schwarz sector. 
\begin{table}\begin{center}\begin{tabular}{llll}\hline 
${\rm Point\ group}$& $(v_2,v_3,v_4)$& $(w_2,w_3,w_4)$&$s$\\\hline 
${\bf Z}_2\times{\bf Z}_2$& $(1,0,-1)/2$&$(0,1,-1)/2$&6\\
${\bf Z}_3\times{\bf Z}_3$& $(1,0,-1)/3$&$(0,1,-1)/3$&56\\
${\bf Z}_2\times{\bf Z}_4$& $(1,0,-1)/2$&$(0,1,-1)/4$&42\\
${\bf Z}_4\times{\bf Z}_4$& $(1,0,-1)/4$&$(0,1,-1)/4$&210\\
${\bf Z}_2\times{\bf Z}_6-I$&$(1,0,-1)/2$&$(0,1,-1)/6$&102\\
${\bf Z}_2\times{\bf Z}_6-II$&$(1,0,-1)/2$&$(1,1,-2)/6$&134\\
${\bf Z}_3\times{\bf Z}_6$& $(1,0,-1)/3$&$(0,1,-1)/6$&272\\
${\bf Z}_6\times{\bf Z}_6$& $(1,0,-1)/6$&$(0,1,-1)/6$&1190\\
\hline\end{tabular}\end{center}\caption{Point-group generators for
 ${\bf Z}_N\times{\bf Z}_M$ orbifolds}\label{ZNxZMorbiflods}\end{table}
For the models of tables \ref{ZNorbiflods} and \ref{ZNxZMorbiflods}, we
 therefore find the following partition function for a flat background 
 generalizing the ${\bf Z}_2\times{\bf Z}_2$ result of \cite{KiritsisKounnas95}
 (for ${\bf Z}_N$ orbifolds set $M=1$):
\ba
&&Z(\tau,\bar\tau)=\frac{1}{4NM}\frac{1}{{\rm Im}\tau|\eta|^4}
\sum_{\alpha,\beta,\bar\alpha,\bar\beta=0}^1\sum_{h_1,g_1=0}^{N-1}
\sum_{h_2,g_2=0}^{M-1}(-)^{\alpha+\beta+\alpha\beta}(-)^{\bar\alpha+
\bar\beta+\bar\alpha\bar\beta}\no\\&&\times Z_2\left[\hspace*{-6pt}
\begin{array}{c}h_1v_2+h_2w_2\\g_1v_2+g_2w_2\end{array}\hspace*{-6pt}
\right]Z_3\left[\hspace*{-6pt}\begin{array}{c}h_1v_3+h_2w_3\\g_1v_3+g_2w_3
\end{array}\hspace*{-6pt}\right]Z_4\left[\hspace*{-6pt}\begin{array}{c}h_1v_4
+h_2w_4\\g_1v_4+g_2w_4\end{array}\hspace*{-6pt}\right]\no\\
&&\times\frac{\theta\left[\hspace*{-6pt}\begin{array}{c}\alpha/2\\
\beta/2\end{array}\hspace*{-6pt}\right]}{\eta}\frac{\theta\left[
\hspace*{-6pt}\begin{array}{c}\alpha/2+h_1v_2+h_2w_2\\\beta/2
+g_1v_2+g_2w_2\end{array}\hspace*{-6pt}\right]}{\eta}\frac{\theta\left[
\hspace*{-6pt}\begin{array}{c}\alpha/2+h_1v_3+h_2w_3\\\beta/2
+g_1v_3+g_2w_3\end{array}\hspace*{-6pt}\right]}{\eta}\frac{\theta\left[
\hspace*{-6pt}\begin{array}{c}\alpha/2+h_1v_4+h_2w_4\\\beta/2
+g_1v_4+g_2w_4\end{array}\hspace*{-6pt}\right]}{\eta}\no\\
&&\times\frac{\bar\theta\left[\hspace*{-6pt}\begin{array}{c}\bar\alpha/2\\
\bar\beta/2\end{array}\hspace*{-6pt}\right]}{\bar\eta}\frac{\bar
\theta\left[\hspace*{-6pt}\begin{array}{c}\bar\alpha/2+h_1v_2+h_2w_2\\
\bar\beta/2+g_1v_2+g_2w_2\end{array}\hspace*{-6pt}\right]}{\bar\eta}
\frac{\bar\theta\left[\hspace*{-6pt}\begin{array}{c}\bar\alpha/2
+h_1v_3+h_2w_3\\\bar\beta/2+g_1v_3+g_2w_3\end{array}\hspace*{-6pt}
\right]}{\bar\eta}\frac{\bar\theta\left[\hspace*{-6pt}\begin{array}{c}
\bar\alpha/2+h_1v_4+h_2w_4\\\bar\beta/2+g_1v_4+g_2w_4\end{array}
\hspace*{-6pt}\right]}{\bar\eta},\no\\&&\label{orbifoldpartitionfunction}
\ea
where $Z_i\left[\hspace*{-6pt}\begin{array}{c}h/2\\g/2\end{array}
\hspace*{-6pt}\right]$ is the partition function of the complex bosons $Z^i$
 with twists $(h,g)$. We have
\be
Z_i\left[\hspace*{-6pt}\begin{array}{c}0\\0\end{array}\hspace*{-6pt}\right]=
\frac{\Gamma(2,2)}{|\eta|^4},
\ee
where $\Gamma(2,2)$ is the (2,2) lattice sum, and
\be
Z_i\left[\hspace*{-6pt}\begin{array}{c}\frac{h}{2}\\\frac{g}{2}\end{array}
\hspace*{-6pt}\right]=\left|\eta(\tau)\left[\theta\left[\hspace*{-6pt}
\begin{array}{c}\frac{1-h}{2}\\\frac{1-g}{2}\end{array}\hspace*{-6pt}\right]
(0,\tau)\right]^{-1}\right|^2\label{twistedbosoncontribution}
\ee
for $(h,g)\not=(0,0)$ (see e.g. \cite{Polchinski}). We introduce a constant
 background curvature parametrized by ${\cal R}$ by the perturbation 
\be
\int d^2z\,{\cal R}[I^3+:\!\psi^1\psi^2\!:][\bar I^3+\!:\tilde\psi^1\tilde
\psi^2\!:]
\ee
of the sigma model action, where $I^i=k{\rm Tr}[\sigma^i g^{-1}\partial g]$
 and $g=\exp[i\sigma\cdot x/2]$. Let $Q$ be the momentum lattice associated
 to the left-moving $U(1)$ current \mbox{:$\psi^1\psi^2$:\ ,} $I$ the charge
 lattice of the left-moving $U(1)$ current associated to $I_3$ and let
 $\bar Q$ and $\bar I$ be defined in terms of the right-moving currents. For a
 pair of dimensions the 2 dimensional Majorana--Weyl world-sheet spinors are
 given by one complex fermion
\be
\psi=\frac{1}{\sqrt{2}}(\psi_1+i\psi_2),\qquad \bar\psi=\frac{1}{\sqrt{2}}(
\psi_1-i\psi_2).
\ee
The fermions can be bozonized (see \cite{Polchinski}) by $\psi=e^{iH}$ and $
\bar\psi=e^{-iH}$ in such a way that $i\partial H=\,:\!\!\psi\bar\psi\!\!:$ .
 The operator $Q$ acts as the fermion-number operator. The undeformed partition
 function can be written as
\be
Z={\rm Tr}\left(q^{L_0}\bar q^{\bar L_0}\right),\qquad q=e^{2\pi i\tau}
\ee
where
\be
L_0=\frac{1}{2}Q^2+\frac{I^2}{k}+\cdots,\qquad \bar L_0=\frac{1}{2}\bar Q^2
+\frac{\bar I^2}{k}+\cdots,
\ee
where the dots stand for operators that do not involve $I, \bar I, Q, \bar Q$.
 In the heterotic string theory the constant background field ${\cal R}$
 transforms $L_0$ and $\bar L_0$ to
\ba
L_0'-L_0=\bar L_0'-\bar L_0=(Q+I)+\frac{\sqrt{1+k(k+2){\cal R}^2}-1}{2}
\left[\frac{(Q+I)^2}{k+2}+\frac{I^2}{k}\right]^2.
\ea
Let us expand the partition function in a power series in ${\cal R}$
\be
Z({\cal R})=\sum_{n=0}^{\infty}{\cal R}^n Z_n.
\ee
In string theory the partition function is already the generating functional
 of the connected Green functions (this is in contrast to field theory where
 one would have to take the logarithm) and the one-loop correction to the
 Einstein--Hilbert action in the effective action is therefore given by
\be
Z^{{\rm het}}_1=-4\pi{\rm Im}\tau\langle(Q+I)\bar I\rangle.
\ee
As $\langle \bar I\rangle=0$ on any genus Riemann surface, we find that the
 Planck mass is not renormalized in perturbation theory for heterotic
 backgrounds with ${\cal N}\geq1$ space-time supersymmetry (see also
 \cite{KiritsisKounnas95} and \cite{Minahan88} to
 \cite{ForgerOvrutTheisenWaldram96}). In type II string theory the
 renormalization of the Planck mass is given by
\be
Z^{{\rm II}}_1=-2\pi{\rm Im}\tau\langle(Q+I)(\bar Q+\bar I)\rangle.
\label{renormalizationofthePlanckmassalaKiritsisKounnas}
\ee
The contribution of the pair of (transverse) world-sheet fermions that
 correspond to the 4-dimensional space to the partition function is given
 by (see \cite{Polchinski})
\be
Z^{\alpha}_{\beta}(\tau)={\rm Tr}_{\alpha}\left[q^He^{i\pi Q\beta}\right]=
\frac{\theta\left[\hspace*{-6pt}\begin{array}{c}\alpha/2\\\beta/2\end{array}
\hspace*{-6pt}\right](0,\tau)}{\eta(\tau)}.
\ee
This was also used to derive (\ref{orbifoldpartitionfunction}). A similar
 computation leads to
\be
{\rm Tr}_{\alpha}\left[q^He^{i\pi Q\beta}Q\right]=\frac{1}{\eta(\tau)}
\frac{1}{2\pi i}\left.\frac{\partial}{\partial_\nu}\theta\left[\hspace*{-6pt}
\begin{array}{c}\alpha/2\\\beta/2\end{array}\hspace*{-6pt}\right](\nu,\tau)
\right|_{\nu=0}.\label{replacingtheta11byitsderivative}
\ee
To compute (\ref{renormalizationofthePlanckmassalaKiritsisKounnas}) we
 therefore can simply replace the corresponding left and right theta functions
 in (\ref{orbifoldpartitionfunction}) by their derivative. We can use the
 following Riemann identity
\ba
&&\frac{1}{2}\sum_{a,b=0}^1(-)^{\alpha+\beta+\alpha\beta}\theta\left[
\hspace*{-6pt}\begin{array}{c}\alpha/2\\\beta/2\end{array}
\hspace*{-6pt}\right](\nu,\tau)\theta\left[\hspace*{-6pt}\begin{array}{c}
\frac{\alpha+h_1}{2}\\\frac{\beta+g_1}{2}\end{array}\hspace*{-6pt}\right]
(0,\tau)\theta\left[\hspace*{-6pt}\begin{array}{c}\frac{\alpha+h_2}{2}\\
\frac{\beta+g_2}{2}\end{array}\hspace*{-6pt}\right](0,\tau)\theta\left[
\hspace*{-6pt}\begin{array}{c}\frac{\alpha-h_1-h_2}{2}\\\frac{\beta-g_1-g_2}
{2}\end{array}\hspace*{-6pt}\right](0,\tau)\no\\&=&\theta\left[\hspace*{-6pt}
\begin{array}{c}1/2\\1/2\end{array}\hspace*{-6pt}\right]
(\nu/2,\tau)\theta\left[\hspace*{-6pt}\begin{array}{c}\frac{1-h_1}{2}\\
\frac{1-g_1}{2}\end{array}\hspace*{-6pt}\right](\nu/2,\tau)\theta\left[
\hspace*{-6pt}\begin{array}{c}\frac{1-h_2}{2}\\\frac{1-g_2}{2}\end{array}
\hspace*{-6pt}\right](\nu/2,\tau)\theta\left[\hspace*{-6pt}\begin{array}{c}
\frac{1+h_1+h_2}{2}\\\frac{1+g_1+g_2}{2}\end{array}\hspace*{-6pt}\right]
(\nu/2,\tau).\label{Riemannidentityfororbifolds} 
\ea
For $\nu=0$ this shows that the partition function
 (\ref{orbifoldpartitionfunction}) vanishes and that we have at least one
 unbroken supersymmetry. We find
\ba
&-&2\pi{\rm Im}\tau\langle Q\bar Q\rangle=\frac{1}{NM}\frac{-2\pi
{\rm Im}\tau}{{\rm Im}\tau|\eta|^4}\sum_{h_1,g_1=0}^{N-1}
\sum_{h_2,g_2=0}^{M-1}\frac{1}{2\pi i\eta^4}\frac{1}{2\pi i\bar\eta^4}\no\\
&\times& Z_2\left[\hspace*{-6pt}\begin{array}{c}h_1v_2+h_2w_2\\g_1v_2+g_2w_2
\end{array}\hspace*{-6pt}\right]Z_3\left[\hspace*{-6pt}\begin{array}{c}h_1v_3
+h_2w_3\\g_1v_3+g_2w_3\end{array}\hspace*{-6pt}\right]Z_4\left[\hspace*{-6pt}
\begin{array}{c}h_1v_4+h_2w_4\\g_1v_4+g_2w_4\end{array}\hspace*{-6pt}\right]
\no\\
&\times&\frac{\partial}{\partial\nu}\Bigl[\theta\left[\hspace*{-6pt}
\begin{array}{c}1/2\\1/2\end{array}\hspace*{-6pt}\right]
\left(\frac{\nu}{2},\tau\right)\theta\left[\hspace*{-6pt}\begin{array}{c}
1/2-h_1v_2-h_2w_2\\1/2-g_1v_2-g_2w_2\end{array}\hspace*{-6pt}
\right]\left(\frac{\nu}{2},\tau\right)\no\\&&\qquad \theta\left[\hspace*{-6pt}
\begin{array}{c}1/2-h_1v_3-h_2w_3\\1/2-g_1v_3-g_2w_3
\end{array}\hspace*{-6pt}\right]\left(\frac{\nu}{2},\tau\right)\left.\theta
\left[\hspace*{-6pt}\begin{array}{c}1/2-h_1v_4-h_2w_4\\1/2
-g_1v_4-g_2w_4\end{array}\hspace*{-6pt}\right]\left(\frac{\nu}{2},\tau\right)
\Bigr]\right|_{\nu=0}\no\\
&\times&\frac{\partial}{\partial\bar\nu}\Bigl[\bar\theta\left[\hspace*{-6pt}
\begin{array}{c}1/2\\1/2\end{array}\hspace*{-6pt}\right]
\left(\frac{\bar\nu}{2},\tau\right)\bar\theta\left[\hspace*{-6pt}
\begin{array}{c}1/2-h_1v_2-h_2w_2\\1/2-g_1v_2-g_2w_2
\end{array}\hspace*{-6pt}\right]\left(\frac{\bar\nu}{2},\tau\right)\no\\&&
\qquad \bar\theta\left[\hspace*{-6pt}\begin{array}{c}1/2-h_1v_3-h_2w_3
\\1/2-g_1v_3-g_2w_3\end{array}\hspace*{-6pt}\right]\left(
\frac{\bar\nu}{2},\tau\right)\left.\bar\theta\left[\hspace*{-6pt}\begin{array}
{c}1/2-h_1v_4-h_2w_4\\1/2-g_1v_4-g_2w_4\end{array}
\hspace*{-6pt}\right]\left(\frac{\bar\nu}{2},\tau\right)\Bigr]
\right|_{\bar\nu=0}.\no\\\label{orbifoldrenormalizationofPlanckmass}
\ea
Because of 
\be
\theta\left[\hspace*{-6pt}\begin{array}{c}1/2\\1/2\end{array}
\hspace*{-6pt}\right](0,\tau)=0\label{theta11iszero},
\ee
the partial derivatives with respect to $\nu$ and $\bar\nu$ have to act on
 the first theta functions. From the $N^2M^2$ possible sectors, those do
 not contribute where either
\ba
h_1v_2+h_2w_2=g_1v_2+g_2w_2&=&0\qquad {\rm mod 1}\\
{\rm or}\qquad h_1v_3+h_2w_3=g_1v_3+g_2w_3&=&0\qquad {\rm mod 1}\\
{\rm or}\qquad h_1v_4+h_2w_4=g_1v_4+g_2w_4&=&0\qquad {\rm mod 1}.
\ea
All sectors that contribute give the same contribution. This is because they
 are all twisted, and in this case the contribution of the twisted bosons
 cancels the one of the twisted fermions (see
 (\ref{twistedbosoncontribution})). Let $s$ be the number of contributing
 sectors. The different values of $s$ are summarized in tables
 \ref{ZNorbiflods} and \ref{ZNxZMorbiflods}. We are left with
\ba
-2\pi{\rm Im}\tau\langle Q\bar Q\rangle&=&\frac{s}{NM}\frac{-2\pi
{\rm Im}\tau}{{\rm Im}\tau|\eta|^4}\frac{1}{2\pi i\eta}\frac{\partial}
{\partial\nu}\left.\theta\left[\hspace*{-6pt}\begin{array}{c}1/2\\
1/2\end{array}\hspace*{-6pt}\right](\nu/2,\tau)\right|_{\nu=0}\no\\&&
\times\frac{1}{2\pi i\bar\eta}\frac{\partial}{\partial\bar\nu}\left.\bar
\theta\left[\hspace*{-6pt}\begin{array}{c}1/2\\1/2\end{array}
\hspace*{-6pt}\right](\bar\nu/2,\tau)\right|_{\bar\nu=0}.
\ea
Using
\be
\left.\partial_{\nu}\theta\left[\hspace*{-6pt}\begin{array}{c}1/2\\
1/2\end{array}\hspace*{-6pt}\right](\nu,\tau)\right|_{\nu=0}=-2\pi
\eta(\tau)^3,\qquad \frac{\partial}{\partial\nu}=\frac{1}{2}\frac{\partial}
{\partial\frac{\nu}{2}},
\ee
we arrive at
\be
-2\pi{\rm Im}\tau\langle Q\bar Q\rangle=s\frac{\pi}{2NM}.
\ee
Multiplying with the fundamental domain integral
\be
\int\limits_{{\cal F}}\!\!\frac{d^2\tau}{({\rm Im}\tau)^2}=\frac{\pi}{3}
\ee
we find the one-loop renormalization of the Planck mass for type II string
 theory compactified on symmetric orbifolds
\be
\Delta{\cal L}_{{\rm eff}}=s\frac{\pi^2}{6NM}M_s^2\sqrt{g}R+\cdots,
\label{LeffZ2xZ2}
\ee
where the dots include higher derivative terms and we have recovered the
 string scale $M_{\rm s}$.\\

In the case of the ${\bf Z}_2$ orbifold we have $(v_2,v_3,v_4)=
\frac{1}{2}(1,0,-1)$ and therefore four supersymmetries. There are then two
 vanishing theta functions of the type (\ref{theta11iszero}) in
 (\ref{orbifoldrenormalizationofPlanckmass}) and the derivative can only act
 on one of them. There are therefore, as expected, no sectors that contribute,
 and the one-loop renormalization of the Planck mass vanishes.\\

With the values of the GUT scale $M_{{\rm \small{GUT}}}\approx 2\times
 10^{16}$ GeV, the Planck mass $M_{\rm P}\approx 10^{19}$ GeV and
 $16\pi\approx 50$ (see (\ref{EinsteinHilbert})) we find
\be
\frac{M_{\rm P}^2}{16\pi}\approx 5000 M_{{\rm \small{GUT}}}^2.
\ee
In the case of the ${\bf Z}_6\times{\bf Z}_6$ orbifold we find the
 one-loop renormalization of the Einstein--Hilbert action of
\be
\frac{1190\pi^2}{6^3}M_{\rm s}^2\approx 50M_{\rm s}^2.
\ee
Under the assumption that the total value of the Planck mass is entirely due
 to one loop (i.e. the tree value is small, e.g. of the order of the GUT scale
 or vanishing) this would give a string scale of
\be
M_{\rm s}\approx 2\times 10^{17}{\rm GeV}\approx 10M_{{\rm \small{GUT}}}.
\ee
Amusingly, this value of $M_{\rm s}/M_{{\rm \small{GUT}}}$ 
is close to the one obtained at tree level
using the experimental value of $\alpha_{\rm \small{GUT}}$.

According to the argument put forward in \cite{GV} a  value for 
$M_{\rm s}/M_{{\rm \small{GUT}}}$ closer to 1 could result from a theory
possessing a large number $N$ of species (``flavours"). In order to
find a superstring model containing an arbitrarily large $N$,
let us consider type IIB string theory compactified on $M_4\times
 EH_3$, where $M_4$ is 4-dimensional Minkowski space and $EH_3$ is the
 Eguchi--Hanson space (see \cite{Polchinski}) that is a Calabi--Yau space with
 3 complex, i.e. 6 real, dimensions and is the same as the orbifold
 $T_6/{\bf Z}_3$ where the singularities have been blown up. As in
 \cite{DiVecciaLerdaMerlatti02} we consider a background of $N_b$ coincident
 D5-branes wrapping a supersymmetric 2-cycle of $EH_3$. The total one-loop
 renormalization of the Planck mass is given by
\be
\Delta{\cal L}_{{\rm \small{eff}}}=\left(\frac{4\pi^2}{9}+c N_b\right)
M_s^2\sqrt{g} R+\cdots,
\ee
where the first term is (\ref{LeffZ2xZ2}) with $s=8$ for the ${\bf Z}_3$
 orbifold and the second term is the contribution of the open strings that
 end on the branes that is determined by the constant $c$. At low energies
 the massless open string modes that can propagate lead to pure ${\cal N}=1$
 super Yang--Mills with gauge group $SU(N_b)$ in 4 dimensions (see
 \cite{DiVecciaLerdaMerlatti02}) with gauge coupling (see \cite{DiVeccia99}) 
\be
g_{YM}^2=2g_s(2\pi)^{p-2}(\alpha')^{(p-3)/2},\qquad p=5.
\ee
The field theory computation using the heat kernel regularization with the
 string scale $M_{\rm s}$ as cut-off yields, in leading order in the coupling
 constant, the contribution of $N_b$ ${\cal N}=1$ vector multiplets
\be
\frac{N_b}{64\pi^2} M_s^2\sqrt{g} R\qquad {\rm or} \qquad c=\frac{1}{64\pi^2}.
\ee
If we assume that the string scale is of the order of the GUT scale and that
 the large value of the Planck mass is entirely due to one loop, this leads to
 $N_b\approx 3\times 10^6$. It does definitely not seem natural to have a
 configuration of such a large number of coincident branes as the vacuum
 state of string theory. The problem is that we have only considered the
 leading contribution at weak coupling for the massless spectrum 
(gauge fields) of the open strings that end on the branes, and not the whole
 tower of states. A full string computation may eventually give much larger
 contributions to the one-loop renormalization of the Planck mass (using the
 same type of field-theory computation, the result that corresponds to
 (\ref{LeffZ2xZ2}) is also much smaller than the full string result), but it
 will still have a term proportional to $N_b M_s^2\sqrt{g} R$. The needed
 number of branes may correspondingly be smaller.

\section{Conclusion}
 By considering orbifold compactifications of type II string theory we have
 shown that a rather large one-loop renormalization of the Planck mass is
 possible, depending on the choice of  compactification within the set that
 preserves ${\cal N}=2$ space-time supersymmetry in 4 dimensions. This is in
 contrast to heterotic string models, where there is no renormalization for
 any compactification that preserves ${\cal N}\geq1$ space-time supersymmetry
 in 4 dimensions. As discussed in the introduction, in order to be able to
 lower the string scale towards the GUT scale, a large number $N$ entering in
 a different way gauge and gravity loops is needed.  Unfortunately, string
 backgrounds containing such a free parameter must involve brane
 configurations for which adequate loop techniques are still lacking. We hope,
 however, to have shown that a parametrically large renormalization of the
 Planck mass (in string units) is all but impossible.
\section*{Acknowledgements}

I thank Gabriele Veneziano for suggesting this problem and for many useful
 discussions and ideas.


\begin{thebibliography}{99}
\bibitem{Dienes96} K.R. Dienes, hep-th/9602045, Phys. Rept. {\bf 287} (1997)
 447.
\bibitem{GV} G. Veneziano, hep-th/0110129, JHEP {\bf 06} (2002) 051.
\bibitem{KiritsisKounnas95} E. Kiritsis and C. Kounnas, hep-th/9501020, Nucl.
 Phys. {\bf B442} (1995) 472.
\bibitem{KiritsisKounnas94} E. Kiritsis and C. Kounnas, hep-th/9410212, Nucl.
 Phys. Proc. Suppl. {\bf 41} (1995) 331.
\bibitem{KiritsisKounnas95a} E. Kiritsis and C. Kounnas, hep-th/9507051, in
 Proc. String 95, Future Perspectives in String Theory, Los Angeles, CA, 1995.
\bibitem{KiritsisKounnas95b} E. Kiritsis and C. Kounnas, hep-th/9509043, in
 Proc. Conf. on Gauge Theories, Applied Supersymmetry and Quantum Gravity,
 Leuven, 1995.
\bibitem{PetropoulosRizos96} P.M. Petropoulos and J. Rizos, hep-th/9601037,
 Phys. Lett. {\bf B374} (1996) 49.
\bibitem{KiritsisKounnasPetropoulosRizos96} E. Kiritsis, C. Kounnas, P.M.
 Petropoulos and J. Rizos, hep-th/9605011, in Proc. $5^{th}$ Hellenic School
 and Workshop on Elementary Particle Physics, Corfu, Greece, 1995.
\bibitem{Petropoulos96} P.M. Petropoulos, hep-th/9605012, in same Proc. as
 Ref. \cite{KiritsisKounnasPetropoulosRizos96}.
\bibitem{KiritsisKounnasPetropoulosRizos96a} E. Kiritsis, C. Kounnas, P.M.
 Petropoulos and J. Rizos, hep-th/9608034, Nucl. Phys. {\bf B483} (1997) 141.
\bibitem{Polchinski} J. Polchinski, {\it String theory volume I and II}
 (Cambridge University Press, 1998).
\bibitem{BailinLove99} D. Bailin and A. Love, Phys. Rep. {\bf 315} (1999)
 285.
\bibitem{Minahan88} J.A. Minahan, Nucl. Phys. {\bf B298} (1988) 36.
\bibitem{AntoniadisGavaNarain92} I. Antoniadis, E. Gava and K.S. Narain, Phys.
 Lett. {\bf B283} (1992) 209.
\bibitem{ForgerOvrutTheisenWaldram96} K. F\"orger, B.A. Ovrut, S.J. Theisen
 and D. Waldram, hep-th/9605145, Phys. Lett. {\bf B388} (1996) 512.
\bibitem{DiVecciaLerdaMerlatti02} P. Di Vecchia, A. Lerda and P. Merlatti,
 hep-th/0205204.
\bibitem{DiVeccia99} P. Di Vecchia, hep-th/9903007, Fortsch. Phys. {\bf 48}
 (2000) 87.


\end{thebibliography}
\end{document}